# Towards an AI Observatory for the Nuclear Sector: A tool for anticipatory governance

Aditi Verma
University of Michigan
Ann Arbor, Michigan, USA
aditive@umich.edu

Elizabeth Williams
The Australian National University
Canberra, Australia
elizabeth.williams@anu.edu.au

## Abstract

AI models are rapidly becoming embedded in all aspects of nuclear energy research and work but the safety, security, and safeguards consequences of this embedding are not well understood. In this paper, we call for the creation of an anticipatory system of governance for AI in the nuclear sector as well as the creation of a global AI observatory as a means for operationalizing anticipatory governance. The paper explores the contours of the nuclear AI observatory and an anticipatory system of governance by drawing on work in science and technology studies, public policy, and foresight studies.

## 1 Introduction

For many countries, nuclear energy is seen as a key ingredient of low-carbon energy systems - a view that is contributing to significant growth in the nuclear sector worldwide. At the same time, a circular relationship between AI and nuclear energy is rapidly emerging in which AI energy demand is driving interest in new nuclear energy technologies and simultaneously AI models are themselves becoming rapidly embedded in all aspects of research and work in the nuclear sector. Though largely 'safe', nuclear energy technologies are nevertheless high-hazard technologies, and the embedding of AI in nuclear work and research, including its implications for safety, security, and safeguards, is not well understood.

The safe operation of the existing reactor designs is the result of systems of safety and governance that are decades in the making and include extensive trial and error by organizations across the nuclear industry [31, 32, 48], information sharing, training, and accreditation of operators and other professionals in the nuclear sector [33, 40], creation of norms and a culture of safety [19, 26], the development of predictive models for understanding and mitigating modes of failure in reactor systems [47], and extensive oversight and governance by national regulatory agencies as well as, in some cases, self-regulators [40]. The emergence and rapid embedding of AI in the nuclear sector (particularly at a time when new nuclear energy technologies are approaching deployment) has the potential to undermine the existing nuclear safety infrastructure by introducing new and unexpected modes of failure in existing and new reactor systems, creating unanticipated security and safeguards vulnerabilities, and deskilling operational, regulatory, and even research personnel. As noted, the potential implications of embedding AI in the nuclear sector are not well understood and only now are starting to be articulated by nuclear safety regulators around the world [8]. Given these concerns, and the need to steward a safe embedding of AI in the nuclear sector even as we potentially transition to use new nuclear energy technologies, we propose to create an anticipatory system of governance for AI in the nuclear sector. This paper outlines our reasoning for proposing this approach to support appropriate use of AI in the nuclear sector.

Many of the existing regulatory approaches for the nuclear sector stem from lessons learned from past nuclear accidents, such as Three Mile Island, Chernobyl, and Fukushima. Studies of such accidents have repeatedly demonstrated that complex and sometimes unpredictable or emergent combinations of technical, human, and organizational interactions and failures [10] often play a role (see [51] for a brief summary of several accidents and subsequent findings). History has taught us the impact of such accidents (even with no lives lost) is global and can include large scale displacements of humans, disruption of economic activity, impacts on energy prices and energy security, cancellation of nuclear construction projects, and reputational losses for the nuclear industry [23]. It is for this reason that an oft-repeated saying in the nuclear industry is that "an accident anywhere is an accident everywhere" [2]. The nuclear power industry (and therefore, the energy industry as a whole) ebbs and flows with societal trust in nuclear energy systems; the vitality of the nuclear power industry is marked by periods of de-investment that are directly timed to nuclear accident-related headlines.

The decisions on how the technical, human, and organizational elements of nuclear systems interact, and the institutional and normative environments in which they function, are made on different timescales, in many cases even decades before an accident occurs (a phenomenon referred to as "slow disasters" which result from poor or deferred decision-making [25]). At the Fukushima Daiichi site, for example, decisions about the height of the seawall, the placement of the emergency diesel generators, the design of the containment (and many other factors) were made before construction began, but all ultimately contributed to the severity of the accident. Decisions being made now concerning the integration of AI in the design, operation, and regulation of nuclear energy technologies, if not made carefully and in a well-considered manner, could potentially be contributory or even initiating factors to accidents in the future as well as security and safeguards vulnerabilities. All of this is to say that the consequences of AI adoption in nuclear energy systems, and the potential interaction of AI models with each other and with the human, organizational, institutional, and technical parts of the system have to be *anticipated*.

Anticipatory governance, which has its roots in science and technology studies [17], administration and management [7], and environmental studies and policy [15], can be defined as the societal capacity to sense, synthesize, and act on a diverse set of signals, trends, and drivers to manage an emerging technology [17, 44]. This form of governance is proposed as a means of governing emergent technologies characterized as being both "high stakes and high uncertainty" for their likely use cases, and deliberately integrates local contexts, diverse knowledge systems, and interactions between social and technical systems of relevance to the technology being governed in its process [42].

The integration of AI in nuclear facilities represents the integration of two classes of technologies that may in their own right meet the "high stakes and high uncertainty" characteristics mentioned above. Appropriately identifying how to predict and manage future problems resulting from the integration of AI in nuclear contexts is therefore urgent — particularly when considering the possible significant expansion of the use of nuclear energy around the world driven, in a circular way, in large part by the energy needs of AI models themselves. At the last two COP conferences over 30 countries committed to tripling nuclear capacity by 2050 [1, 14] (over and above the 400+ nuclear reactors that operate around the world today and the 60+ reactors under construction). An anticipatory system of governance is needed to navigate the potential risks and uncertainties of large-scale use and adoption of AI across an expanding nuclear sector. To demonstrate this, we will: discuss how AI is already being used or researched in the sector; define and explain the safety, security, and safeguards implications such work poses; briefly discuss how AI regulation in safety-critical sectors (including the nuclear industry) are being discussed; and present our argument and approach for employing an anticipatory governance approach — to be supported by an AI observatory similar to that proposed in [20] for genome editing.

## 2 AI in the Nuclear Sector

The adoption of AI in the nuclear sector is proceeding at a rapid pace and largely in the absence of any frameworks or processes of governance [27]. Though the necessity of AI governance in the nuclear sector is recognized by nuclear safety regulators [8], few concrete measures have yet to be taken to create such governance.

To our knowledge, the most significant guidance on AI governance in the sector comes from the US Nuclear Regulatory Commission (NRC), which has created an artificial intelligence strategic plan [27] that establishes a typology of automation and potential implications of regulatory oversight. In its strategic plan, the NRC recognizes the need to improve its own readiness for regulatory decision-making, creating an organizational framework for reviewing AI applications, building an AI-proficient workforce as well as building external partnerships.

Careful steps forward are crucial; while guidance from other safety-critical sectors exists, past nuclear accidents have already demonstrated the gravity of any failures in the industry, even in cases with no lives lost [23, 37]. Comparably, the bar is significantly higher in the nuclear industry than it is in other industries, which has contributed to the conservative and highly regulated nature of the industry to date.

The regulatory discussion reflects broader discussions about the applications of AI in the field. While AI has been identified as important for helping the industry modernize [18], several historical, technical, and business barriers to AI adoption in the nuclear sector exist [18, 24] that must be addressed for the industry to move forward.

This is due in large part to the slow pace of regulation in the nuclear sector as well as the ever-evolving and changing landscape of possible applications of AI in nuclear engineering research, reactor design, operations, and maintenance. AI models are being applied, for example, in materials discovery and simulation of materials performance in radiation environments [12, 34]; the design and optimization of many reactor systems and components, including reactor cores [4, 38, 45]; development of schemas for autonomous reactor operation with significant reduction or elimination entirely of human operators [43]; and planning of maintenance and downtime for operating reactors [16]. Not included in this list is the increasing use of generative AI by researchers, practitioners, and students in their daily work (such as the analysis of data, writing or summarization of technical reports, etc) in ways that are not well-documented or understood but which may have significant consequences for nuclear energy systems indirectly by influencing the rigor and attention to detail. ChatOPG (co-developed by Ontario Power Generation and Microsoft — as discussed in [53]) is an example of how such tools are being adopted in the industry. Many universities (including the University of Michigan) have created bespoke generative AI tools that are frequently used by students and faculty alike as part of their daily work.

The rapid embedding of AI in research and work in the nuclear sector has, in the absence of governance, the potential to compromise the *safety*, *security*, and *safeguards* ('3S') of these technologies, as defined below. We will discuss the '3S' factors and their importance, which we will then draw on to explain the need to create an adaptive, anticipatory form of governance for the sector.

## 3 Safety, Security, and Safeguards

In the nuclear industry, the potential impact of any AI-enabled addition to a nuclear facility can be assessed in terms of the technology's likely impact on the '3S' considerations – safety, security, and safeguards. Drawing from [3, 52], *Safety* focuses on ensuring that operating conditions prevent accidents, mitigate accident consequences, and protect workers, the public and the environment from the risks of unintentional releases of radiation; *Security* refers to protections designed to prevent intentional radiological releases, theft through unauthorized access to nuclear facilities; and *Safeguards* is the prevention, primarily through materials accountancy, of (explosive) military use of nuclear technologies, either by states or terrorist organizations.

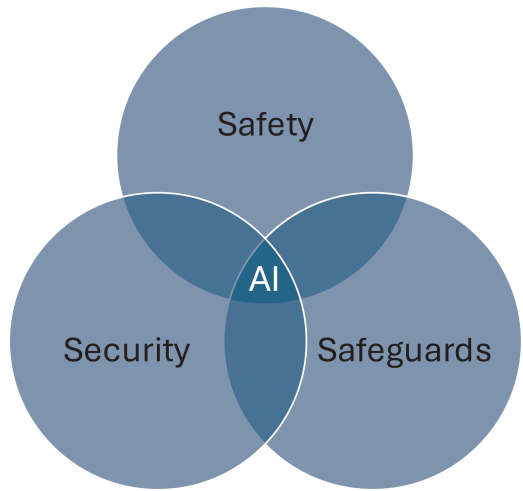

**Figure 1: Adapted from [52]. Nuclear technologies are complex, as are many AI technologies. Integrating AI therefore requires careful consideration of '3S' impacts.**

These '3S' considerations can be thought of as distinct values that must be upheld in the nuclear industry to support continued operation, and are often regulated by different local, state, national, and even international bodies. However, the '3S's are interlinked, sometimes in complex and even contradictory ways [52] — as illustrated in Fig. 1. This makes managing such challenges an act of constant vigilance — and as argued in [52], suggests a need for an "all-hazards" definition of safety in the sector.

We placed AI at the center of Fig. 1 because some forms of AI add complexity to a system, and also may bring in an additional temporal dimension (through training data or design decisions) that are not always straightforward to manage. Sub-symbolic AI, for instance, describes a diverse suite of computational models that use statistical approaches to learning based on data and encompasses approaches such as reinforcement learning, artificial neural networks, swarm intelligence, and self-organizing maps. In nuclear systems, the use of such tools in a range of applications (as direct or indirect integrations within the broader nuclear 'system') can make '3S' challenges harder to identify and manage. This suggests the AI model creation pipeline should be considered in assessments of '3S' impacts across the nuclear facility lifecycle, because everything from the data used to train the models, to the quality control and assurance approaches used to verify the tool or technology can shape AI performance in nuclear contexts.

While formal verification methods for forms of AI exist, these methods have limitations [28]. The extent to which AI is being integrated into such facilities – e.g. to reduce human workforce requirements [18] or create digital twins [46] – mean that some applications may not lend themselves well to formal verification. In particular, human verification or oversight is often used to ensure appropriate performance of such systems [8], which means human-factor considerations and questions around how to foster and maintain appropriate trust in such AI integrations shape whether verifiable AI systems appropriately fulfill their intended purpose in practice. AI systems that impact '3S' considerations in complex ways can ease workloads, but can also add to the already significant toll on human workers in the nuclear industry, particularly in light of the fact that such workers have long been considered an essential safety system in nuclear plants [37].

### 3.1 The relationship between nuclear systems, other safety-critical fields, and existing AI regulatory approaches

The nuclear industry is one of a number of highly-regulated "high-hazard" industries (e.g. aviation, oil and gas, medicine, defense) that are grappling with whether, when and how to use AI in their sectors. These industries each have distinct regulatory approaches that predate the introduction of AI, and those pre-existing structures tend to influence how AI governance is being enacted. However, there are some shared approaches that are relevant across some industries. In particular, safety cases — which in part classify plant components according to their relevance to safety, provide an extended analysis of a given plant design as relevant to regulatory requirements, and are meant to discuss a licensing discussion with a regulator — are used in nuclear, automotive, and aviation industries. Limited examples of work designed to address AI using safety case analyses for such sectors exist in the literature (e.g. [5, 41]). Such work is very limited in scope, is challenging to extend, and in both of the cited examples, is based on hypothetical AI-enabled systems. It also parallels a discussion (coming out of the UK AI Safety Institute) about extending safety case analyses to 'frontier AI' (see [6]). This literature leaves unanswered questions, and suggests the level of certainty required for nuclear applications in particular will pose a challenge using traditional safety case approaches.

## 4 What is anticipatory governance?

An anticipatory system of governance treats the technological system in question as a sociotechnical system, and has participatory and futuring elements to it that are designed to make room for contribution from diverse perspectives and knowledge systems and plan across long timescales. The approach consists of five main components: hindsight, insight, topsight, prescience, and engagement [13, 17], each of which are defined as follows: (1) *Hindsight* calls for a historically grounded understanding of the emerging technology being considered, including an awareness and knowledge of analogous technologies and their governance, which may offer a set of 'better practices' to emulate; (2) *Insight* is a knowledge of the intentions, biases, and assumptions of the actors seeking to govern; (3) *Topsight* is a knowledge of how the emerging technology being considered interacts with and is enmeshed with other technologies and their systems of governance; and (4) *Prescience* can be thought of as an "attunement to weak signals that faintly hint at what is possible" [13] and in the context of an emerging technology such as AI (specifically in the nuclear sector), the weak signals that point to possible futures can take the form of extrapolations from active areas of research and inquiry to their potential applications in industry. This leads to (5) *engagement*, which calls for understanding not just the expert imaginaries for an emerging technology but also more widely held sociotechnical imaginaries [21] including specifically the hopes, fears, and concerns surrounding an emerging technology across different publics [30].

Anticipatory governance has been applied or has been attempted in a number of circumstances, including national security and intelligence[11], climate change [39], and a broad range of emerging technologies including self-driving cars, solar geoengineering, and carbon sequestration [22]. The latter set of emerging technology examples have not created *entire* systems of governance but have instead, in a one-off manner, sought to create the public engagement component of the governance system through participatory technology assessments. Significant variation across applications of anticipatory governance is also seen in the intent behind creating the system of governance [35]:

*Risk reduction:* Some efforts may be motivated by scoping possible or even plausible futures to carry out strategic planning, reduce risks, and navigate uncertainties.

*Democratic futures and co-creation*: Another possibility is that anticipatory governance may be used to create more pluralistic futures through societal mobilization and co-creation of alternatives (relative to the extant trajectory).

*Interrogating visions of preferred futures*: Yet another possibility is that anticipatory governance may be used to uncover how sociotechnical imaginaries (collective visions of preferred futures) are effectively 'performing' such futures. These imaginaries and the way they are performed [29], once uncovered, can be interrogated and corrected.

Different actors and organizations participating in an anticipatory system of governance may be motivated by different considerations. For the specific case of building an anticipatory system of governance for AI in the nuclear sector, the aspiration should be to build a system responsive to each intent, but especially the latter two as the current trajectory of AI in the nuclear sector is purely technocratic and is almost certainly driven by a vision of nuclear facilities that no longer require a human safety component.

## 4.1 Why Anticipatory Governance of AI is needed in the Nuclear Sector

As noted above, anticipatory governance of AI is needed in the nuclear sector because of the far-reaching implications on safety, security, and safeguards, particularly when the potential risks and uncertainties of large-scale use and adoption of AI across an expanding nuclear sector are considered. Of equal importance is the need to interrogate where, how, and why AI is being used across the nuclear energy technology life cycle, how much of that use (or expected use) is driven by behaviors shaped by preferred future visions, and to explore whether a more socially informed, co-creative assessment and adoption of AI is possible. These concerns, though enumerated in the context of the nuclear sector, apply perhaps broadly to a range of technology sectors using AI (and perhaps even to the development of AI as a whole).

## 5 Towards a framework of anticipatory governance for AI

What might an anticipatory form of governance for AI for the nuclear sector look like?

As noted, hindsight, insight, topsight, prescience, and engagement can be thought of as the building blocks of a system of anticipatory governance. Some preliminary features of each of these sub-systems in the context of the nuclear sector are described below:

**Hindsight**: The history of regulation in many sectors, including in the nuclear sector, has been marked by regulatory capture —- occurring as a result of insufficient organizational, regulatory, financial, and legal separation of the regulator from those who seek to develop and promote the technology or sector in question [9, 50]. Given the significant negative consequences of maladaptation of AI in the nuclear sector, great care must be taken to prevent capture of an anticipatory system of governance while ensuring those who govern are equipped with the tools, skills, and people to do so. Notably, in the nuclear sector, it is important to understand *who* possesses skills at the intersection of nuclear science and engineering and AI, and should gaps or skill asymmetries exist, reserves of these skills must be built within independent systems of governance. The distribution of these skills as well as gaps are not well documented.

**Insight**: Insight is needed in how AI is being used across the nuclear technology sector in the first place. As described at the start of this paper, the uses of AI in nuclear work are many, ever expanding, and not well-documented. At a minimum, declarations of AI use must be required not only as part of the peer-review and publication process (as is now standard practice) but also across every aspect of work in the nuclear sector. These declarations are a starting point. In addition to or as part of the declarations, it is crucial to understand the intention and motivation behind embedding AI in nuclear work. This work can be done by creating a nuclear AI observatory (or more broadly, an observatory [20] of AI embedding in high hazard and consequence technology sectors).

**Topsight:** Nuclear technologies are complex systems in and of themselves and they are enmeshed with still other complex systems such as the power grid (and soon potentially coupled directly with industrial users). Across these levels of sociotechnical systems and sub-systems, it is important to anticipate and explore how AI integration alters existing or creates new risks and modes of failure and build in failsafes in safety-critical systems in particular. Eliminating or significantly reducing human presence in these systems, while reducing the potential for human error, also reduces the possibility of human improvisation and ingenuity which have proven to be significant factors in preventing initiating events from snowballing into accidents and safety improvements overall [36].

**Prescience**: Beyond anticipating risks and modes of failures resulting from AI integration in nuclear work, it is important to also anticipate how the integration of AI in the nuclear sector alters the very nature of nuclear work and which futures we are working towards. For example, it is important to understand how the integration of AI in nuclear energy research, design, regulation, and operations changes the daily work of nuclear professionals, their engagements with each other, and the extent to which they verify the outputs of AI models they or others work with.

*Engagement:* Most importantly, each aspect of the system of anticipatory governance must be informed by the values of the publics (plural). What preferences, hopes, fears, and concerns does the public have concerning the embedding of AI in the nuclear sector? These perspectives can be investigated through an ongoing program of sociotechnical assessment [49] and co-creation with publics.

# 6 Establishing an observatory and anticipatory system of governance for AI in the nuclear sector

In the next phase of this work, we seek to create an AI nuclear observatory to seek answers to the questions underlined above as well as to observe and assess the implementation of anticipatory forms of governance in the nuclear sector on a small scale across several sites where nuclear work is done before larger, more systemic implementation.

This work of creating the nuclear AI observatory will begin with a documentation of the uses of AI in the nuclear sector through an annual survey which will be disseminated internationally to researchers, practitioners, regulators, and students. We acknowledge that novel applications of AI are emerging and may be considered proprietary by their developers. Where this is the case, we will ask respondents to describe their inventions in broad terms wherever possible. We will solicit interview participants through the survey as well as our respective professional networks in the US and in Australia. The survey data and analyses of the interview findings will be hosted on a website that will be the digital home of the nuclear AI observatory, which, taking inspiration from the Global Observatory on Gene Editing, will serve as a universal clearinghouse for information on AI in the nuclear sector. In addition to data and information on AI adoption in the nuclear sector, the Observatory will also host resources on AI and nuclear energy, intended for a general audience. We will also regularly monitor the regulatory and policy developments on AI in the nuclear sector and post these on the observatory website along with opinion, analysis, and news pieces. We will hold periodic workshops to bring together AI experts, nuclear experts, and members of the public. The purpose of these workshops will be to elicit perspectives from each group on the potential benefits, risks, and harms arising from the embedding of AI in the nuclear sector. Each of these workshops will also initiate preliminary discussions on the question of norms and practices governing AI adoption, and more broadly, the contours of an anticipatory system of governance. The final set of workshops will focus on these questions in detail with the aim of generating consensus-based norms for AI adoption and measures for operationalizing anticipatory governance. Recognizing, as noted above, that not all participants will be able to openly share their work on AI in the nuclear sector, we will explore the creation of processes (including a confidential review board) that provide recommendations and produce summaries of trends for AI uses that cannot be disclosed publicly.

# References

[1] [n. d.]. U.S. Sets Targets to Triple Nuclear Energy Capacity by 2050. https://www.energy.gov/ne/articles/us-sets-targets-triple-nuclear-energy-capacity-2050. Accessed: 2025-3-25.

[2] 2005. The Enduring Lessons of Chernobyl. https://www.iaea.org/newscenter/statements/enduring-lessons-chernobyl. Accessed: 2025-4-11.

[3] 2022. *IAEA Nuclear Safety and Security Glossary.* INTERNATIONAL ATOMIC ENERGY AGENCY, Vienna. https://www.iaea.org/publications/15236/iaea-nuclear-safety-and-security-glossary Book.

[4] Rabie Abu Saleem, Majdi I Radaideh, and Tomasz Kozlowski. 2020. Application of deep neural networks for high-dimensional large BWR core neutronics. *Nucl. Eng. Technol.* 52, 12 (Dec. 2020), 2709–2716.

[5] Christopher R. Anderson and Louise A. Dennis. 2023. Autonomous Systems' Safety Cases for use in UK Nuclear Environments. In *AREA@ECAI*. https://api.semanticscholar.org/CorpusID:263322549

[6] Marie Davidsen Buhl, Gaurav Sett, Leonie Koessler, Jonas Schuett, and Markus Anderljung. 2024. Safety cases for frontier AI. arXiv:2410.21572 [cs.CY] https://arxiv.org/abs/2410.21572

[7] Günter Bächler. 2004. Conflict transformation through state reform. In *Transforming Ethnopolitical Conflict*. VS Verlag für Sozialwissenschaften, Wiesbaden, 273–294.

[8] Canadian Nuclear Safety Commission, UK Office for Nuclear Regulation, US Nuclear Regulatory Commission. 2024. CONSIDERATIONS FOR DEVELOPING ARTIFICIAL INTELLIGENCE SYSTEMS IN NUCLEAR APPLICATIONS.

[9] Daniel Carpenter and David A Moss. 2013. *Preventing regulatory capture: Special interest influence and how to limit it.* Cambridge University Press, Cambridge, England.

[10] Sidney Dekker, Paul Cilliers, and Jan-Hendrik Hofmeyr. 2011. The complexity of failure: Implications of complexity theory for safety investigations. *Saf. Sci.* 49, 6 (July 2011), 939–945.

[11] L Fahey and R M Randall. 1997. *Learning from the future: Competitive foresight scenarios.* John Wiley & Sons, Nashville, TN.

[12] K Field, R Jacobs, Mingen Shen, Matthew E Lynch, Priyam V Patki, C Field, and D Morgan. 2021. Development and deployment of automated machine learning detection in electron microcopy experiments. *Microsc Microanal* 27, S1 (July 2021), 2136–2137.

[13] Leon S Fuerth. 2009. Foresight and anticipatory governance. *Foresight* 11, 4 (July 2009), 14–32.

[14] Jenny Gross. 2023. 22 Countries Pledge to Triple Nuclear Capacity in Push to Cut Fossil Fuels. *The New York Times* (Dec. 2023).

[15] Aarti Gupta. 2002. Searching for shared norms : global governance of biosafety. *Unpublished Doctoral Dissertation, Yale University. October* (2002).

[16] Ezgi Gursel, Bhavya Reddy, Anahita Khojandi, Mahboubeh Madadi, Jamie Baalis Coble, Vivek Agarwal, Vaibhav Yadav, and Ronald L. Boring. 2023. Using artificial intelligence to detect human errors in nuclear power plants: A case in operation and maintenance. *Nuclear Engineering and Technology* 55, 2 (Feb. 2023), 603–622. doi:10.1016/j.net.2022.10.032

[17] David H Guston. 2014. Understanding 'anticipatory governance'. *Soc. Stud. Sci.* 44, 2 (April 2014), 218–242.

[18] Anna Hall and Vivek Agarwal. 2024. Barriers to adopting artificial intelligence and machine learning technologies in nuclear power. *Progress in Nuclear Energy* 175 (Oct. 2024), 105295. doi:10.1016/j.pnucene.2024.105295

[19] Institute of Nuclear Power Operations. 2012. *Traits of a Healthy Nuclear Safety Culture.* Technical Report.

[20] Sheila Jasanoff and J Benjamin Hurlbut. 2018. A global observatory for gene editing. *Nature* 555, 7697 (March 2018), 435–437.

[21] Sheila Jasanoff and Sang-Hyun Kim. 2015. *Dreamscapes of Modernity: Sociotechnical Imaginaries and the Fabrication of Power.* University of Chicago Press.

[22] Leah R Kaplan, Mahmud Farooque, Daniel Sarewitz, and David Tomblin. 2021. Designing Participatory Technology Assessments: A Reflexive Method for Advancing the Public Role in Science Policy Decision-making. *Technol. Forecast. Soc. Change* 171 (Oct. 2021), 120974.

[23] Ioannis N. Kessides. 2012. The future of the nuclear industry reconsidered: Risks, uncertainties, and continued promise. *Energy Policy* 48 (2012), 185–208. doi:10.1016/j.enpol.2012.05.008 Special Section: Frontiers of Sustainability.

[24] Khalid, Muhammad Hammad, Bui, Ha, Farshadmanesh, Pegah, Al Rashdan, Ahmad, and Mohaghegh, Zahra. 2023. Automation Trustworthiness in Nuclear Power Plants: A Literature Review. https://www.ideals.illinois.edu/items/129041

[25] S Knowles. 2014. Learning from disaster?: The history of technology and the future of disaster research. *Technology and Culture* 55 (Dec. 2014), 773–784.

[26] C Linde. 2005. Shouldering risks: The culture of control in the nuclear power industry. *Technology and Culture* (2005).

[27] M. Dennis, T. Lalain, L. Betancourt, A. Hathaway, and R. Anzalone. 2023. *Artificial Intelligence Strategic Plan: Fiscal Years 2023-2027.* Letter report NUREG-2261. US Nuclear Regulatory Commission, Washington DC, USA.

[28] M. L. Bolton, E. J. Bass, and R. I. Siminiceanu. 2013. Using Formal Verification to Evaluate Human-Automation Interaction: A Review. *IEEE Transactions on*

*Systems, Man, and Cybernetics: Systems* 43, 3 (May 2013), 488–503. doi:10.1109/TSMCA.2012.2210406

[29] D MacKenzie. 2008. *An Engine, Not a Camera: How Financial Models Shape Markets*. MIT Press.

[30] Maureen McNeil. 2013. Between a rock and a hard place: The deficit model, the diffusion model and publics in STS. *Sci. Cult. (Lond.)* 22, 4 (Dec. 2013), 589–608.

[31] R Meserve. 2009. The global nuclear safety regime. *Daedalus* 138 (Sept. 2009), 100–111.

[32] R Meserve. 2022. Strengthening the global nuclear safety regime. *Nuclear Law* (2022).

[33] Joseph S Miller, Bob Stakenborghs, and Robert W Tsai. 2012. Improving nuclear power plant's operational efficiencies in the USA. In *Volume 1: Plant Operations, Maintenance, Engineering, Modifications, Life Cycle, and Balance of Plant; Component Reliability and Materials Issues; Steam Generator Technology Applications and Innovatio*. ASME, 1–8.

[34] Dane Morgan, Ghanshyam Pilania, Adrien Couet, Blas P Uberuaga, Cheng Sun, and Ju Li. 2022. Machine learning in nuclear materials research. *Curr. Opin. Solid State Mater. Sci.* 26, 2 (April 2022), 100975.

[35] Karlijn Muiderman, Aarti Gupta, Joost Vervoort, and Frank Biermann. 2020. Four approaches to anticipatory climate governance: Different conceptions of the future and implications for the present. *Wiley Interdiscip. Rev. Clim. Change* 11, 6 (Nov. 2020).

[36] Akiyoshi Obonai, Takao Watanabe, and Kazuo Hirata. 2014. Successful cold shutdown of onagawa: The closest nuclear power station to the march 11, 2011, epicenter. *Nucl. Technol.* 186, 2 (May 2014), 280–294.

[37] President's Commission on the Accident at Three Mile Island. 1979. *Report of the President's Commission on the accident at Three Mile Island. The need for change: the legacy of TMI*. Technical Report. President's Commission on the Accident at Three Mile Island, Washington, DC (United States). doi:10.2172/6986994

[38] Emilian Popov, Richard Archibald, Briana Hiscox, and Vladimir Sobes. 2022. Artificial intelligence-driven thermal design for additively manufactured reactor cores. *Nuclear Engineering and Design* 395 (Aug. 2022), 111862. doi:10.1016/j.nucengdes.2022.111862

[39] Ray Quay. 2010. Anticipatory governance: A tool for climate change adaptation. *J. Am. Plann. Assoc.* 76, 4 (Sept. 2010), 496–511.

[40] Joseph V Rees. 2009. *Hostages of each other: The transformation of nuclear safety since Three Mile Island*. University of Chicago Press, Chicago, IL.

[41] Alexander Rudolph, Stefan Voget, and Jürgen Mottok. 2018. A consistent safety case argumentation for artificial intelligence in safety related automotive systems. In *9th European Congress on Embedded Real Time Software and Systems (ERTS 2018) (9th European Congress on Embedded Real Time Software and Systems (ERTS 2018))*. Toulouse, France. https://hal.science/hal-02156048

[42] {R. W. Foley} and {D. H. Guston} and {D. Sarewitz}. 2019. Towards the Anticipatory Governance of Geoengineering. In *Geoengineering Our Climate?* (1 ed.). Routledge, 223–243. https://doi.org/10.4324/9780203485262-40

[43] Hanan A Saeed, Minjun Peng, Hang Wang, and Athar Rasool. 2023. Autonomous control model for emergency operation of small modular reactor. *Annals of Nuclear Energy* 190 (Sept. 2023), 109874. doi:10.1016/j.anucene.2023.109874

[44] Scott Smith and Madeline Ashby. 2020. *How to Future: Leading and Sense-making in an Age of Hyperchange*. Kogan Page Publishers.

[45] Vladimir Sobes, Briana Hiscox, Emilian Popov, Rick Archibald, Cory Hauck, Ben Betzler, and Kurt Terrani. 2021. AI-based design of a nuclear reactor core. *Scientific Reports* 11, 1 (Oct. 2021), 19646. doi:10.1038/s41598-021-98037-1

[46] Vaibhav Yadav, Hongbin Zhang, Christopher P. Chwasz, Andrei V. Gribok, Christopher Ritter, Nancy J. Lybeck, Ross D. Hays, Timothy C. Trask, Prashant K. Jain, Vittorio Badalassi, Pradeep Ramuhalli, Doug Eskins, Ramon L Gascot, Daniel Ju, and Raj Iengar. 2021. *The State of Technology of Application of Digital Twins*. Letter report TLR/RES-DE-REB-2021-01. US Nuclear Regulatory Commission, Washington DC, USA.

[47] Aditi Verma. 2024. A tale of two epistemologies : the evolution of nuclear safety in the US and French nuclear industry. *Entrep. Hist.* 1 (July 2024), 48–69.

[48] Aditi Verma, Ali Ahmad, and Francesca Giovannini. 2021. Nuclear energy, ten years after Fukushima. *Nature* 591, 7849 (March 2021), 199–201.

[49] Aditi Verma and Todd Allen. 2024. A sociotechnical readiness level framework for the development of advanced nuclear technologies. *Nucl. Technol.* (March 2024).

[50] T Wellock. 2012. Engineering uncertainty and bureaucratic crisis at the atomic energy commission, 1964–1973. *Technology and Culture* 53 (Nov. 2012), 846–884.

[51] Ronald J. Willey. 2024. . By , Cambridge, MA . . 9760262546881| 9870262376761 (epub). $75.00 Hardcover,. *Process Safety Progress* 43, 1 (2024), 221–221. doi:10.1002/prs.12573 arXiv:https://aiche.onlinelibrary.wiley.com/doi/pdf/10.1002/prs.12573

[52] Adam D. Williams. 2020. Systems Theory Principles and Complex Systems Engineering Concepts for Protection and Resilience in Critical Infrastructure: Lessons from the Nuclear Sector. *INSIGHT* 23, 2 (June 2020), 14–20. doi:10.1002/inst.12293 Number: 2 Publisher: John Wiley & Sons, Ltd.

[53] Darryl Willis. 2023. The era of AI: Transformative AI solutions powering the energy and Resources Industry. https://www.microsoft.com/en-us/industry/blog/energy-and-resources/2023/09/28/the-era-of-ai-transformative-ai-solutions-powering-the-energy-and-resources-industry/